# Transport in a binary mixture of non-spherical molecules: Is hydrodynamics at fault?

**Sarmistha Sarkar, Tuhin Samanta and Biman Bagchi***

*Solid State and Structural Chemistry Unit, Indian Institute of Science, Bangalore 560012, India.*
*E-mail**: bbagchi@iisc.ac.in; profbiman@gmail.com

*Abstract*

We consider a new class of model systems to study systematically the role of molecular shape in the transport properties of dense liquids. Our model is a liquid binary mixture where both the molecules are non-spherical and characterized by a collection of parameters. Although in the real world most of the molecules are non-spherical, only a limited number of theoretical studies exist on the effects of molecular shapes and hardly any have addressed the validity of the hydrodynamic predictions of rotational and translational diffusion of these shapes in liquids. In this work, we study a model liquid consisting of a mixture of prolate and oblate (80:20 mixture) ellipsoidals with interactions governed by a modified Gay-Berne potential for a particular aspect ratio (ratio of the length and diameter of the ellipsoids), at various temperature and pressure conditions. We report calculations of transport properties of this binary mixture by varying temperature over a wide range, at a fixed pressure. We find that for the pressure-density conditions studied, there is no signature of any phase separation, except transitions to the crystalline phase at low temperatures and relatively low pressure (the reason we largely confined our studies to high pressure). *We find that for our model binary mixture, both the stick and slip hydrodynamic predictions break down in a major fashion, for both prolates and oblates and particularly so for rotation*. Moreover, prolates and oblates themselves display different dynamical features in the mean square displacement and in orientational time correlation functions.

## I. INTRODUCTION

The first expressions of friction on a moving molecule were derived by Maxwell, Boltzmann and Enskog who used kinetic theory of gases to obtain the expressions[1] we find in undergraduate texts. In these studies, friction originates from collisions between molecules resulting in an exchange of momentum. The alternate description of friction ($\zeta$) is offered by Navier-Stokes hydrodynamics[2] where the interaction between the tagged molecule and the surrounding solvent (approximated as a viscous continuum) is described by a set of hydrodynamic conditions. Hydrodynamics offers surprisingly simple expressions of translational and rotational friction ($\zeta$) where the only inputs are the viscosity ($\eta$) of the medium and a factor describing the size (could be radius R for spheres) and shape (the aspect ratio for ellipsoidal) of the tagged molecules. For translational friction, hydrodynamics allows derivation of expressions for the two hydrodynamic boundary conditions, namely, stick and slip.[2] The expressions differ only by a factor of 3/2, for spheres, with the larger value ($6\pi\eta R$) for the stick boundary condition (b.c.). The situation, however, is quite different for a rotation where the two boundary conditions give completely different predictions.

The hydrodynamic description of the situation where a spherical molecule can rotate without experiencing any friction is known as the slip boundary condition limit. On the other hand, when a layer of solvent rotates with the tagged sphere, it is called stick b.c. As expected, rotational friction for slip b.c. is zero while quite large $(8\pi\eta R^3)$ for the stick b.c. One often finds experimentally[3] that stick b.c. over-estimates the value of the friction significantly.

A connection between friction and experimental observables such as translational diffusion is provided by Einstein's relation,[4] $D = \frac{k_B T}{\zeta}$. The same holds for rotation, and is often described as Debye-Einstein relation.

The prediction of zero friction for rotating (with slip) spheres posed a paradoxical situation because at least in theoretical studies we like to model molecules as spheres which serve to obtain a simple theoretical expression. The unphysical large value predicted by the stick b.c. further complicated the issue. The paradoxical situation was first resolved by Hu and Zwanzig.[5] They showed for the first time that a reasonable value of orientational correlation time for the rotation of a probe is obtained if the solute is considered as a spheroidal with slip b.c., which is a more realistic description of the molecules studied in experiments. Furthermore, the motion of different spheroids can be different. For example, the motion of oblate is quite different from that of a prolate.[3, 6-10] Unfortunately, there does not exist sufficient data in the literature to discuss about the shape dependence of diffusion.

While experimental studies always deal with molecules that are of nonspherical shape, theoretical models, particularly in computer simulation studies, often invoke spherical molecules, with only a few exceptions. Although the theoretical choice of spheres is mainly motivated by the possibility of simplified calculations, there now exist a number of hydrodynamic expressions for ellipsoidal molecules in closed forms.[11] The situation is especially unusual when a theory is developed and implemented, based on spherical molecules, such as the mode coupling theory[12] in its traditional form and then used to explain orientational relaxation observed in Nuclear magnetic resonance (NMR) or dielectric relaxation that studies non-spherical molecules.[13, 14]

Diffusion of ellipsoids finds extensive use in the study of biopolymers, like proteins.[11] From the beginning, Perrin'sequations[15] and corrections published later[16] have been used to obtain a measure of the asymmetry in the shape of a protein molecule. The standard practice has been to fit the experimental data on rotational diffusion (extracted from NMR) to fit to Perrin's equations and find the asymmetry parameter κ.

Indeed, very little is known about the molecular shape dependence diffusion of rod-shaped and disc-shaped molecules. While the former can diffuse preferentially in the direction of long-axis,[7,11] the latter is likely to exhibit the saucer-like motion because of the expected tendency of the disc-shaped molecule to translate in the plane of the disc, though tumbling of its axis helps it in changing its direction of motion.

As mentioned, there exist many experimental studies on orientational relaxation[17] (like NMR, Kerr effect, fluorescence depolarization). They all measure the time correlation of the second rank spherical harmonics, $Y_{2m}(\Omega(t))$ where $\Omega(t)$ is the time-dependent orientation of the ellipsoidal molecules.[18,19] In this case, the order parameter is $P_2(\cos\theta(t))$ where $P_2$ is the second rank Legendre polynomial of the angle that the major axis of an ellipsoidal makes with the Z-axis of the laboratory-fixed frame.

Microscopic theoretical approach by Medina *et al.*[20,21] showed a strong coupling between the translational and rotational motions while studying the diffusion of nonspherical molecules in suspension of spheres. A similar kind of coupling was found by Chandra and Bagchi[22-24] in the study of orientational relaxation of dipolar molecules.[25]

There is multiple motivation behind this work.

(i) Given the vast difference between stick and slip b.c. in the case of rotation, it is important to understand the actual values of the rotational diffusion constant in various cases. Similarly, the translational diffusion also needs investigation.

(ii) It is important to understand the translational-rotational coupling of ellipsoidal molecules like prolates and oblates. The diffusion of the prolates or diffusion of spheroids in a liquid of Lennard-Jones spheres was studied earlier. Here we present molecular dynamics simulations of oblates in a liquid of prolates using the commercially available molecular dynamics simulation package Large-scale Atomic/Molecular Massively Parallel Simulator (LAMMPS).[26] We show that the translational diffusion is strongly coupled with the rotational diffusion. We compare the simulation results with the predictions of the hydrodynamic theory.

(iii) Temperature dependence of diffusion constant and of the rotational correlation time are subjects of great interest. In the simulation, we can vary these quantities over a considerable range, and study the temperature dependence in detail.

## II. SYSTEM AND SIMULATION DETAILS

We have carried out molecular dynamics (MD) simulations of translational and rotational motions of the binary mixture consisting of total 256 particles of which 205 are prolates and 51 are oblate ellipsoids (disc-shaped molecules) contained in a cubic box with periodic boundary conditions. We have also considered larger system sizes as discussed later. The interactions between any two ellipsoids with arbitrary orientations is assumed to be represented by a modified Gay-Berne (GB) potential.

The expression for the GB potential is given by,

$$U_{GB} = 4\varepsilon(\hat{r},\hat{u}_1,\hat{u}_2)\left[\left[\frac{\sigma_0}{r-\sigma(\hat{r},\hat{u}_1,\hat{u}_2)+\sigma_0}\right]^{12} - \left[\frac{\sigma_0}{r-\sigma(\hat{r},\hat{u}_1,\hat{u}_2)+\sigma_0}\right]^{6}\right], \quad (1)$$

where $\hat{u}_i$ is the axial vector of the molecule $i$ and $\hat{r}$ is the vector along the intermolecular vector $r = r_2 - r_1$, where $r_1$ and $r_2$ denote the centers of mass of molecule 1 and 2, respectively. $\sigma_0$ is the diameter of the minor axis of the ellipsoid. $\sigma(\hat{r},\hat{u}_1,\hat{u}_2)$ and $\varepsilon(\hat{r},\hat{u}_1,\hat{u}_2)$ are the orientation dependent range and strength parameters, respectively. $\sigma(\hat{r},\hat{u}_1,\hat{u}_2)$ is given by,

$$\sigma(\hat{r},\hat{u}_1,\hat{u}_2) = \sigma_0\left[1 - \frac{1}{2}\chi\left[\frac{(\hat{r}.\hat{u}_1 + \hat{r}.\hat{u}_2)^2}{1+\chi(u_1.u_2)} + \frac{(\hat{r}.\hat{u}_1 - \hat{r}.\hat{u}_2)^2}{1+\chi(u_1.u_2)}\right]\right]^{-1/2}, \quad (2)$$

Here, we have used the well-established Gay-Berne pair potential[27-29] represented by GB $(\kappa,\kappa',\mu,\nu)$. In the GB potential, $\kappa$ defines the aspect ratio, that is the ratio of the semimajor axis (a) to the semiminor axis (b). $\sigma$ and $\varepsilon$ depend on the aspect ratio $\kappa$. $\kappa'$ is the energy anisotropy parameter defined by the ratio of the depth of the minimum of the potential for a pair of molecules aligned parallel in a side-by-side configuration to that in an end-to-end configuration. $\mu$ and $\nu$ are two exponents those are adjustable. It follows that the aspect ratio $\kappa$ provides a measure of the shape anisotropy whereas $\kappa'$ provides a measure in the anisotropy of the well depth which can also be controlled by the two parameters $\mu$ and $\nu$. We have employed the parameterization GB(1.2, 3, 2,1) for prolate and GB(0.8, 3, 2,1) for oblate where 1.2 and 0.8 are the aspect ratio of prolate and oblate respectively. We have set anisotropy parameter $\kappa' = 3$. We have considered two parameters as $\mu = 2$ and $\nu = 1$.

In order to verify the possible system-size dependence of our results, we have also studied systems with 500 and 864 ellipsoids. The results obtained for 500 and 864 particles are found to be similar to 256 particles reported in this study. Additionally, to explore the robustness of our results, we have varied separately the aspect ratio ($\kappa$) and the energy anisotropy parameter ($\kappa'$). All the quantities are given in reduced units, defined in terms of the Gay-Berne potential parameters $\sigma_0$ and $\varepsilon_0$, each of which is taken as unity; length in units of $\sigma_0$, temperature in units of $\varepsilon_0/k_B$, where $k_B$ is the Boltzmann constant and time in units of $\left(m\sigma_0^2/\varepsilon_0\right)^{1/2}$, m is the mass of each ellipsoid molecule. As customary, the mass and the moment of inertia of the ellipsoids are set to unity.

The various energy parameters of the interaction potentials for our study have been chosen as follows: $\varepsilon_{OO} = 0.5$, $\varepsilon_{PP} = 1.0$, $\varepsilon_{PO} = 1.5$. It is to be noted that the choice of the energy parameters for the present binary mixture has been motivated by those of the Kob-Anderson binary mixture.[30]

The equations of motion have been integrated using the Velocity-Verlet algorithm with integration time step of dt = 0.001 in the reduced unit. We initially equilibrate the system in NPT ensemble for $10^5$ steps. Further equilibration is performed in NVT ensemble and finally, the data collection has been executed in the NVT ensemble for $5\times10^7$ steps.

## III. HYDRODYNAMIC PREDICTIONS

Coupling between the translational and rotational diffusion[6] of non-spherical molecules possesses considerable interdisciplinary interests[31] due to its importance in physical (liquid crystals),[32] chemical (micelles), and biological (lipids) systems. However, a consensus regarding an appropriate dynamical model is still lacking. The only source is the solution of Navier-Stokes hydrodynamic equations capable of providing expressions for the friction on prolates and oblates. The hydrodynamic solutions provide two different type of situations that depend on whether the stick or the slip boundary condition is applied, as discussed above.

The difference between the predictions of the stick and the slip boundary condition even for non-spherical molecules is of course not new. Evans *et al*[33] showed that the use of stick hydrodynamic condition can be misleading for diffusion of long ellipsoids (prolates) in a liquid. They proved by numerical solution of the Navier-Stokes hydrodynamic equations using the slip boundary condition that the motion along the parallel direction can be completely decoupled from that of perpendicular direction. The overall translational diffusion process, however, remains isotropic.

In fact, the derivation of hydrodynamic expression for friction requires the solution of Navier-Stokes equation for the flow field formed past the sphere which is made stationary by the change of coordinate system with the origin fixed on the sphere. For a sphere, the calculation of the flow for rotation and translation are done separately.

The situation becomes far more complex when the molecule is asymmetric, like a prolate and oblate. In such a situation, a solution of the Navier-Stokes equation is highly non-trivial. This non-trivial nature is reflected in the complexity of the expressions derived by Perrin.[29] In fact, no analytical solution is available for rotational friction on spheroidals with slip boundary condition.

### A. Translational diffusion for prolates and oblates

With the stick hydrodynamic prediction, it is known that for rod-shaped molecules (prolates), the translational diffusion coefficient in a direction parallel to its major axis $(D_{\parallel})$ is equal to the twice of the translational diffusion coefficient in the perpendicular direction $(D_{\perp})$.[11] The main point here is that orientation can change the direction of diffusion. This introduces a mechanism of translation-rotation coupling not present in spherical molecules.

On the other hand, the slip hydrodynamic theory predicts that there could be decoupling between the perpendicular and parallel motion. The ratio between the diffusion coefficient in the parallel to that in the perpendicular direction approaches the value of aspect ratio $(\kappa)$.[33, 34]

For ellipsoids, the stick hydrodynamic predictions for translational diffusion are known analytically and given by,[35]

$$D_{\parallel} = \frac{k_B T \left[(2a^2 - b^2)S - 2a\right]}{(a^2 - b^2)16\pi\eta},\tag{3}$$

and

$$D_\perp = \frac{k_B T \left[(2a^2 - 3b^2)S + 2a\right]}{(a^2 - b^2) 32\pi\eta}, \tag{4}$$

where *a* and *b* are the lengths of the semi-major and semi-minor axes of the ellipsoid.

For a prolate, S is given by,[15, 16]

$$S(P) = \frac{2}{(a^2 - b^2)^{1/2}} \log \frac{a + (a^2 - b^2)^{1/2}}{b}. \tag{5}$$

For an oblate, S is given by,

$$S(O) = \frac{2}{(b^2 - a^2)^{1/2}} \tan^{-1} \frac{(b^2 - a^2)^{1/2}}{a}. \tag{6}$$

These non-trivial expressions were derived originally by Perrin.[15] They reduce to the expression for the sphere when the aspect ratio approaches unity. In the case of calculation of translational diffusion for slip hydrodynamic predictions, we use the table provided by Evans and coworkers.[33]

### B. Rotational diffusion for prolates and oblates

For stick hydrodynamic boundary condition, the rotational diffusion coefficient for ellipsoidal molecule (prolate or oblate) with aspect ratio $\kappa$ can be written as follows:[14]

$$D_R = \frac{3}{2} \times \frac{\kappa\left[(2\kappa^2 - 1)S - \kappa\right]}{\kappa^4 - 1} D_S, \tag{7}$$

where

$$D_S = \frac{k_B T}{6V\eta}. \tag{8}$$

Here the volume V of the ellipsoid is given by, $V = \frac{4}{3}\pi a^2 b.$

In the case of rotational diffusion, S is defined separately for prolates and oblates.

For prolate, S is given by,[7]

$$S(P) = (\kappa^2 - 1)^{-1/2} \ln\left[\kappa + (\kappa^2 - 1)^{1/2}\right]. \tag{9}$$

For oblate we have a different expression,[7]

$$S(O) = (1 - \kappa^2)^{-1/2} \tan^{-1}\left[\frac{(1-\kappa^2)^{1/2}}{\kappa}\right]. \tag{10}$$

Therefore, the values of rotational diffusion (and friction) are completely determined for the stick b.c. The same is not true for the slip b.c. In fact, the study of rotational relaxation of the non-spherical molecule was long hindered by this lacuna till Hu and Zwanzig[5] published the results for the slip b.c. Unfortunately, however, no analytical expression exists for the slip b.c. We have used the table provided by Hu and Zwanzig.[5]

## IV. RESULTS AND DISCUSSION

### A. Temporal evolution of orientational time correlation functions for prolates and oblates

In order to investigate the orientational dynamics, we have calculated second rank orientational correlation functions with respect to time for different temperatures.

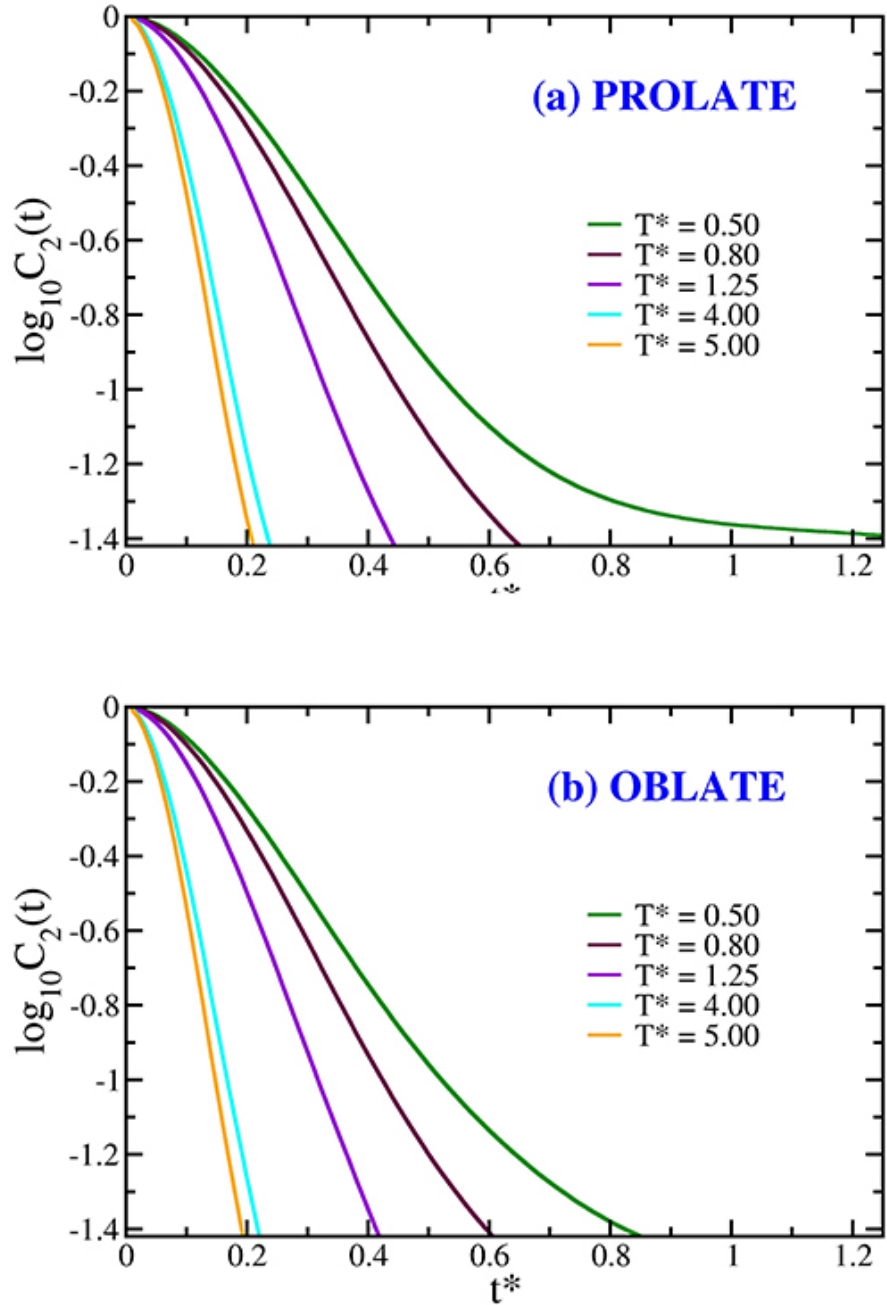

**Figure 1:** Variation of second rank orientational correlation functions with time for (a) prolates and (b) oblates for various temperatures ranging from high T*=5.00 to low-temperature T*=0.50 at pressure P*=30. It is to be noted that the decay of orientational correlation function becomes faster with the increase in temperature. Moreover, the orientational correlation function of prolate ellipsoids shows a slower decay than the oblate, as evident from the plots.

**Figures 1(a)** and **1(b)** display the orientational time-correlation function (OCF) of the second rank spherical harmonics, $C_2(t)$ for prolates and oblates, respectively for different temperatures ranging from high $T^* = 5.00$ to low-temperature $T^*=0.50$ at a constant pressure $P^*=30$. The figures suggest that the decay of orientational relaxation function becomes faster with the increase in temperature. The OCF of prolate ellipsoids shows a slower decay than the oblate, as shown in **Figs. 1(a) and (b)**.

We have also calculated the second rank orientational correlation time $(\tau_2)$ as a function of temperature by integrating the second rank orientational time correlation function. The temperature dependence of the inverse of the second rank orientational correlation time is plotted in **Figures 2** for (a) prolate and (b) oblate.

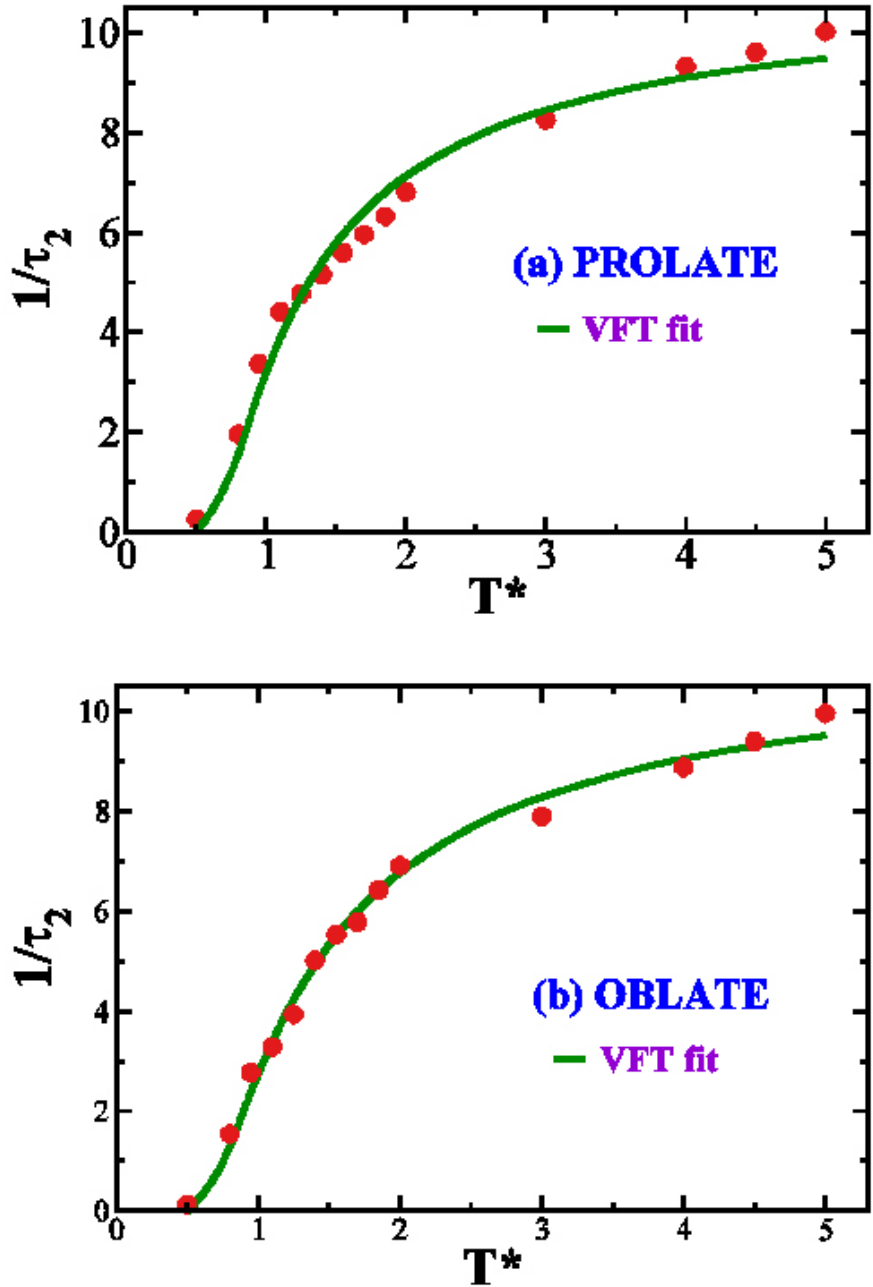

**Figure 2:** The temperature dependence of the inverse second rank orientational correlation time. The plot is shown over the whole temperature range. The green solid lines are the respective VFT fits as shown in figures for (a) prolates and (b) oblates, using the equation, $y(T) = y_0 \exp\left(\dfrac{A}{T-T_g}\right)$. Such fit yields $y_0=10.97$, $A = -0.664$, $T_g = 0.432$ for prolates and $y_0=11.378$, $A = -0.816$, $T_g = 0.435$ for the oblates.

**Figures 2(a) and (b)** show the inverse of $\tau_2$ as a function of temperature over the entire temperature range of our study at a constant pressure P*=30 for prolate and oblate, respectively. The green solid lines are the respective VFT fits for prolates and oblates, as shown in **Fig. 2(a) and Fig. 2(b)**. This VFT fit predicts the glass transition temperature to be $T_g$ = 0.432 for prolates and $T_g$ = 0.435 for the oblates.

Rotational relaxation time is inversely proportional to the volume. We have considered the mass of each prolate and oblate to be equal. This is one of the reasons that we have chosen both the prolate and oblate with same molecular volume. As a result, the relaxation times are quite similar for both prolates and oblates studied here.

### B. Mean square displacement (MSD) of prolate and oblate with time during an increase in temperature: Translational dynamics

In order to focus on the translational dynamics, we calculate the mean squared translational displacements of the center of mass of the prolates as well as oblates. The mean squared displacement of the center of mass is capable of capturing the dynamics of the system with the increase in temperature. At short times, the motion is ballistic and at high temperatures, the ballistic motion is followed by the diffusive motion, as evident in **Fig. 3(a) and (b)**. An intermediate sub-diffusive regime appears just after the ballistic motion, before the onset of the diffusive motion. The occurrence of the intermediate sub-diffusive regime, at low temperatures, arises due to the motion within a cage of the first neighbors.

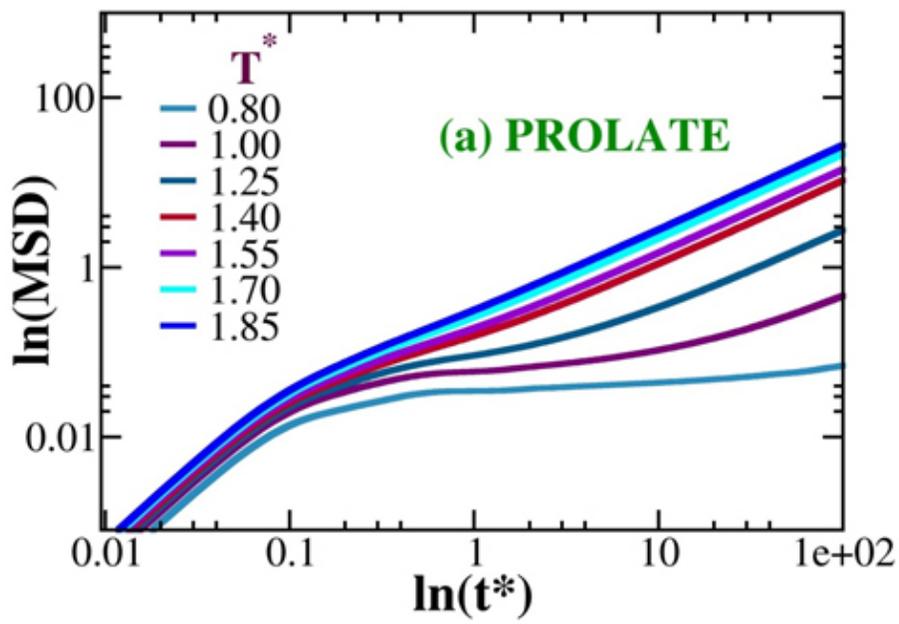

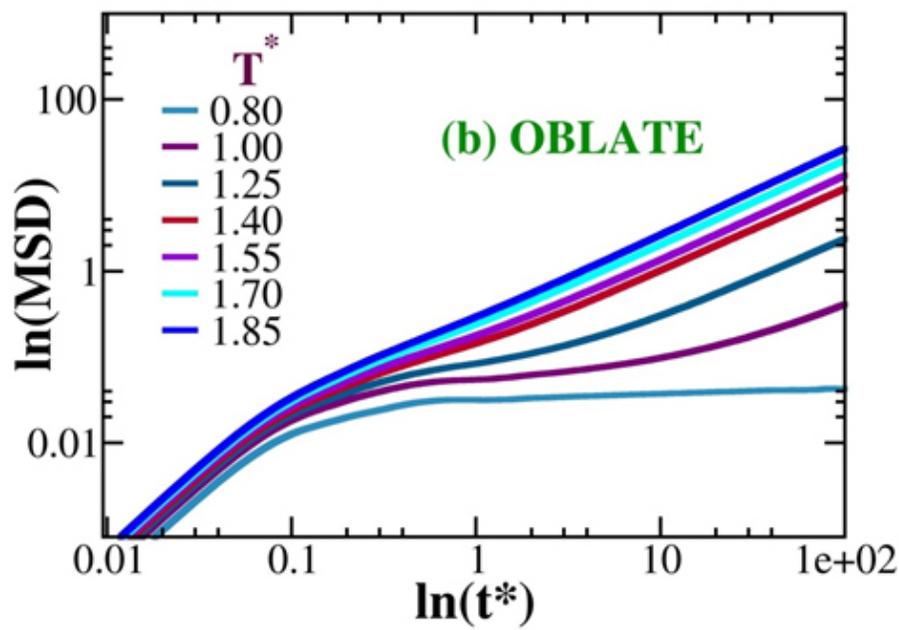

**Figure 3:** Time dependence of the mean square displacement (log-log plot) of the (a) prolate and (b) oblate for a range of temperatures starting from low ($T^* = 0.80$) to high temperature ($T^* = 1.85$).

The temporal variation of diffusivity is another important aspect to understand the translational dynamics of prolates and oblates. In **Fig. 4**, we plot the total diffusion coefficient as a function of temperature.

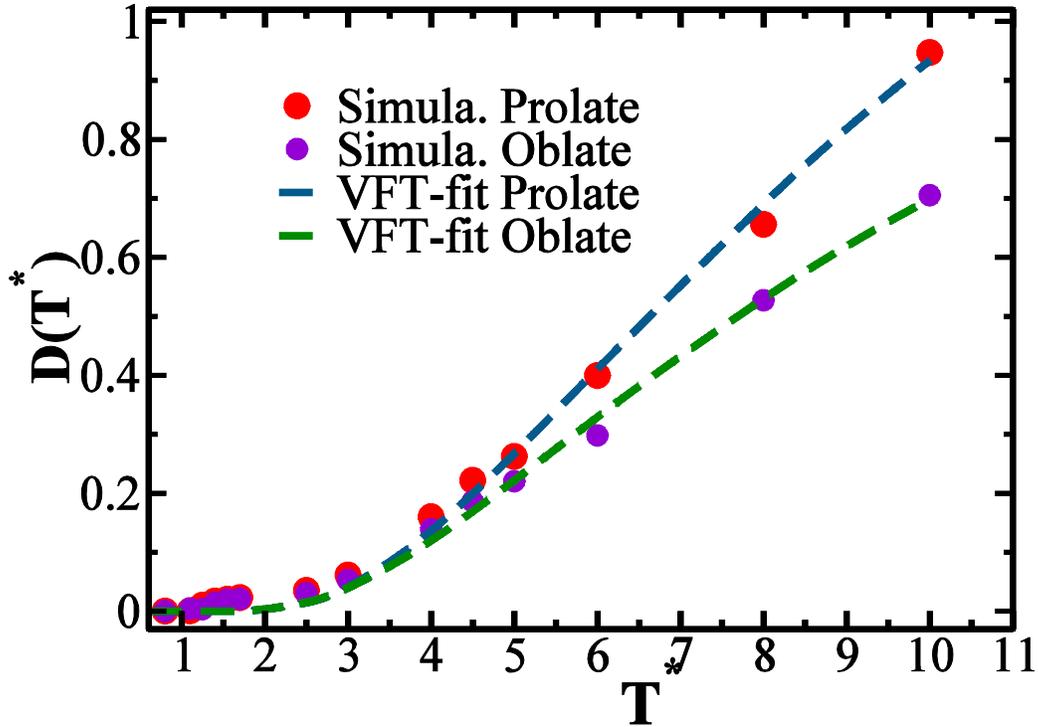

**Figure 4: Plot of the diffusion constant of the system as a function of temperature for prolate and oblate. The red and violet circles correspond to the simulation results for prolates and oblates, respectively. The dashed lines correspond to the VFT fit to the diffusivity data using the equation, $D(T) = D_0 \exp\left(\dfrac{A}{T - T_g}\right)$. The blue and green dashed lines are the VFT fits for prolates and oblates, respectively. Such fits yield $D_0$=2.93, A = -10.94, $T_g$ = 0.43 for prolates and $D_0$=1.99, A = -10.03, $T_g$ = 0.436 for the oblates.**

The diffusivity approaches zero with the decrease in temperature, which seems to follow a non-Arrhenius kind of temperature dependence. We have fitted the calculated values of the diffusion constant to the empirical VFT form, for both prolates and oblates, using the relation $D(T) = D_0 \exp\left(\frac{A}{T - T_g}\right)$. Here $D_0$, $A$, and $T_g$ values are temperature independent parameters. The temperature where the diffusion goes to zero is found to be **0.43** for prolate and **0.436** for oblate. Surprisingly, these values of the apparent glass transition temperature appear to be in good agreement with earlier theoretical results for spheres where the VFT fit gives a value of glass transition temperature close to 0.4. It is to be noted that the value of glass transition temperature ($T_g$) is almost the same for both prolate and oblate. This closeness in the value of $T_g$ arises due to the universality of relaxation phenomena near the glass transition.

### C. Coupling between translational and rotational diffusion

Rotation of solute molecules in a viscous liquid should be coupled to the structural relaxation of the surrounding solvents. Therefore, the rotational diffusion of the solute should be correlated with the translational diffusion coefficient of the surrounding solvent molecules. There is a lack of microscopic theoretical study to interconnect the translational and rotational diffusion of solute molecules. Hence, one has to take the help of hydrodynamics[36] in order to correlate translation and rotation through viscosity.

Here we present the observed inverse relation between orientational correlation time $(\tau_2)$ and the diffusion constant $(D)$ in a viscous liquid for prolates and oblates in terms of slip and stick hydrodynamic predictions.

As expected, the orientational correlation time is directly proportional to the viscosity and the translational diffusion constant of the solvent molecules is inversely proportional to the viscosity. Hence, the frictional resistance acts as the correlating thread between the orientational correlation time and translational diffusion constant. The viscosity dependence of the orientational correlation time comes into play through the viscosity of the medium as rotation in a condensed phase experience friction due to the interaction with the neighboring molecules.

Let us now check the validity of the coupling between translational diffusion and rotational diffusion for slip and stick hydrodynamic predictions. The combination of the Debye-Stokes-Einstein (DSE) and Stokes-Einstein (SE) equations predicts that the product $D\tau_2$ should be independent of temperature even when the shear viscosity increases by many orders of magnitude on approaching the glass transition temperature. **Figure 5** indeed shows such constancy of the product over a wide range of temperatures. Our simulation result also shows that the product of $\tau_2$ and $D$ is only weakly dependent on temperature (and on the viscosity of the liquid) as shown in **figure 5(a)** and **5(b)** for prolate and oblate, respectively, both from simulation results and hydrodynamic predictions.

An earlier molecular dynamics simulation study reported this constancy of the product of diffusion constant and rotational correlation time with respect to the composition of binary mixtures comprising of asymmetric particles.[37]

It is to be noted that the Stokes-Einstein relation works out well when the viscosity is not too high. The SE and DSE relationships together suggest a coupling between translational and rotational diffusion. Such a coupling can be interpreted by the constancy of the product $D\tau_2$. In our case, the simulated value of the ratio of translational and rotational diffusion constant ($D_T/D_R$) is not equal to the value obtained from the hydrodynamic prediction. A recent simulation study[38] has also shown the similar kind of break-down of this expected constancy even for ambient aqueous solutions.

**Figures 5(a) and 5(b)** show the temperature dependence of the quantity $D\tau_2$ for prolate and oblate respectively over the temperature range where the system is in liquid state and the viscosity is comparatively low. This is the reason for the constancy of the product $D\tau_2$ in this temperature range. In several studies, the Stokes-Einstein relationship is found to break down for supercooled liquids[39] at the low-temperature region. We have shown the low-temperature range here (T* = 0.8) in the **Figures 5(a) and (b)** where there is a clear signature of breakdown and it can be explained to arise due to dynamical heterogeneity.

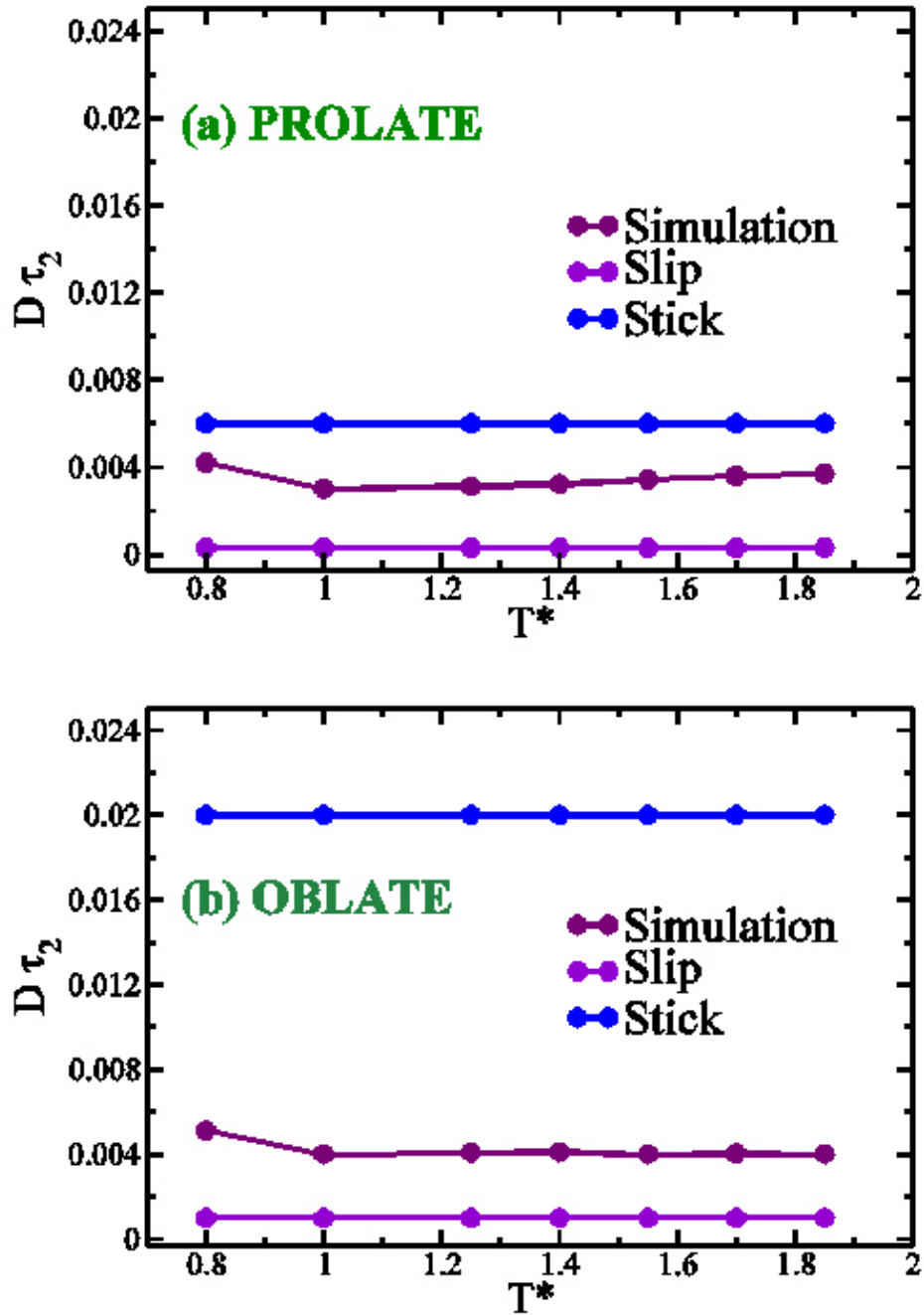

Figure 5: The variation of the product of the translational diffusion coefficient and the second rank orientational correlation time as a function of temperature for the ellipsoids in case of (a) prolate and (b) oblate respectively. It is to be noted that the product $D\tau_2$ should be independent of temperature on approaching the glass transition temperature as predicted by the combination of the Debye-Stokes-Einstein

(DSE) and Stokes-Einstein (SE) equations. Our result is in good agreement showing such constancy of the product $D\tau_2$ over a wide range of temperatures for prolates and oblates both from simulation results and hydrodynamic predictions.

In the following tables, we show the values of $\tau_2$, $D$ and $D\tau_2$ for prolates (Table 1) and oblates (Table 2).

Table 1: We show the values of $\tau_2$, $D$ and $D\tau_2$ for prolates obtained from simulation, slip and stick hydrodynamic b.c. respectively. For our present model binary mixture, both the stick and slip hydrodynamic predictions are found to break down significantly for prolates and in particular for rotation.

| | Table 1: For PROLATES | | | | | | | | |
|---|---|---|---|---|---|---|---|---|---|
| | $\tau_2$ | | | $D$ | | | $D\tau_2$ | | |
| T* | Simu | Slip | Stick | Simu | Slip | Stick | Simu | Slip | Stick |
| 1.85 | 0.165 | 0.019 | 0.63 | 0.024 | 0.015 | 0.0106 | 0.0039 | 0.0003 | 0.0066 |
| 1.55 | 0.176 | 0.023 | 0.773 | 0.022 | 0.012 | 0.0084 | 0.0038 | 0.0003 | 0.0065 |

**Table 2:** We show the values of $\tau_2$, $D$ and $D\tau_2$ for oblates obtained from simulation, slip and stick hydrodynamic b.c. respectively. It is to be noted that for both the stick and slip hydrodynamic predictions, we find a significant breakdown for oblates and also for rotation.

| Table 2: For OBLATES | | | | | | | | | |
|---|---|---|---|---|---|---|---|---|---|
| T* | $\tau_2$ | | | $D$ | | | $D\tau_2$ | | |
| | Simu | Slip | Stick | Simu | Slip | Stick | Simu | Slip | Stick |
| 1.85 | 0.163 | 0.031 | 0.97 | 0.0234 | 0.034 | 0.0212 | 0.0038 | 0.001 | 0.0206 |
| 1.55 | 0.1735 | 0.038 | 1.19 | 0.0216 | 0.028 | 0.0179 | 0.0037 | 0.001 | 0.021 |

*One of the significant results found here [from **figures 5(a) & (b) and tables (1) & (2)**] is that for the present model binary mixture, both the stick and slip hydrodynamic predictions are found to fail considerably both for prolates and oblates, and especially for rotation.*

The product $D\tau_2$ is particularly important as there is no explicit involvement of shear viscosity term. Additionally, the product $D\tau_2$ is appropriate for the comparison with experimental data as $\tau_2$ is directly accessible through experiments.

### D. Crossover from diffusive to jump mode of motion upon cooling

We show here typical trajectories at higher temperature (T* = 1.0) shown in **Fig. 6** for (a) prolate and (b) oblate; at lower temperature (T* = 0.8) for (c) prolate and (d) oblate. There is a significant translational jump[40] at a lower temperature for prolate and oblate, as indicated by **figures 6** (c) and (d), respectively.

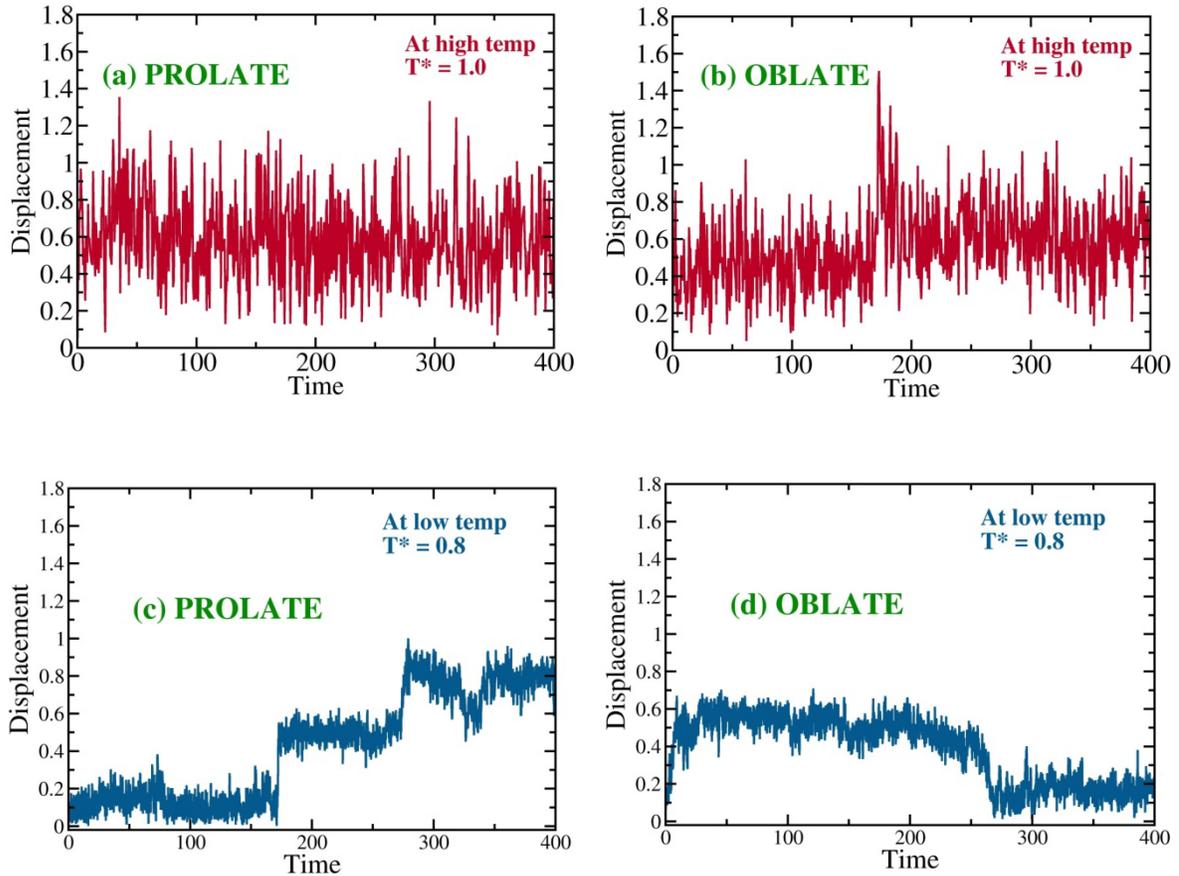

**Figure 6: The plots for the magnitude of the translational displacement vector obtained from the single-particle trajectories for selected ellipsoids over a time window at higher temperature (T* = 1.0) for (a) prolates and (b) oblates; at lower temperature (T* = 0.8) for (c) prolates and (d) oblates are shown. It is to be noted that the signature of translational hops at a lower temperature is evident, as shown in (c) for prolate and (d) for oblate.**

**Figures 7** (a)-(d) show typical trajectories in real orientational space at a higher temperature (T* = 1.0) for (a) prolate and (b) oblate; at lower temperature (T* = 0.8) for (c) prolate and (d) oblate. The occurrence of angular jumps at a lower temperature (T* = 0.8) by an angle close to $180^o$ is prominent, as shown in **Fig. 7** (c) for prolate and (d) for oblate. At low temperature, the rotational jump motions are found to be more prominent than the translational counterparts. This may be due to the head-tail symmetry of revolution for ellipsoidal molecules.

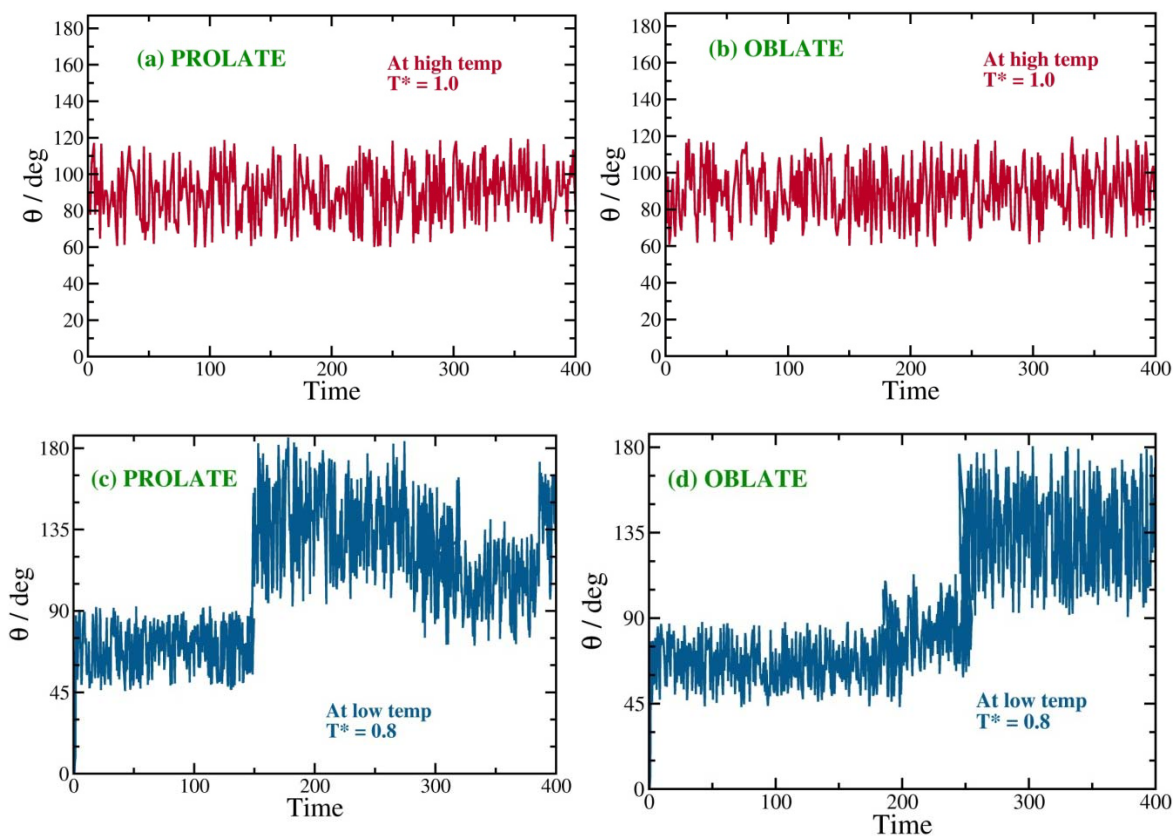

**Figure 7:** The displacement in the polar angle θ obtained from the trajectories for selected ellipsoids over a time window, at higher temperature (T* = 1.0) for (a) prolates and (b) oblates; at lower temperature (T* = 0.8) for (c) prolates and (d) oblates. The occurrence of large angular jumps (by an angle close to $180^o$) at a lower temperature is evident, for (c) prolate and (d) oblate.

### E. Microscopic characterization of the prolate-oblate binary mixture system at a temperature

Partial radial distribution functions for the prolate-oblate binary mixture system at higher (T* = 1.85) and lower temperature (T* = 0.50) are shown in **Figs. 8(a)** and **8(b)** respectively. **Fig. 8(a)** shows that oblate-oblate correlation is the weakest compared to that of prolate-oblate and prolate-prolate correlation. Hence, the probability of finding an oblate surrounded by other oblates is less than the probability of being surrounded by prolates.

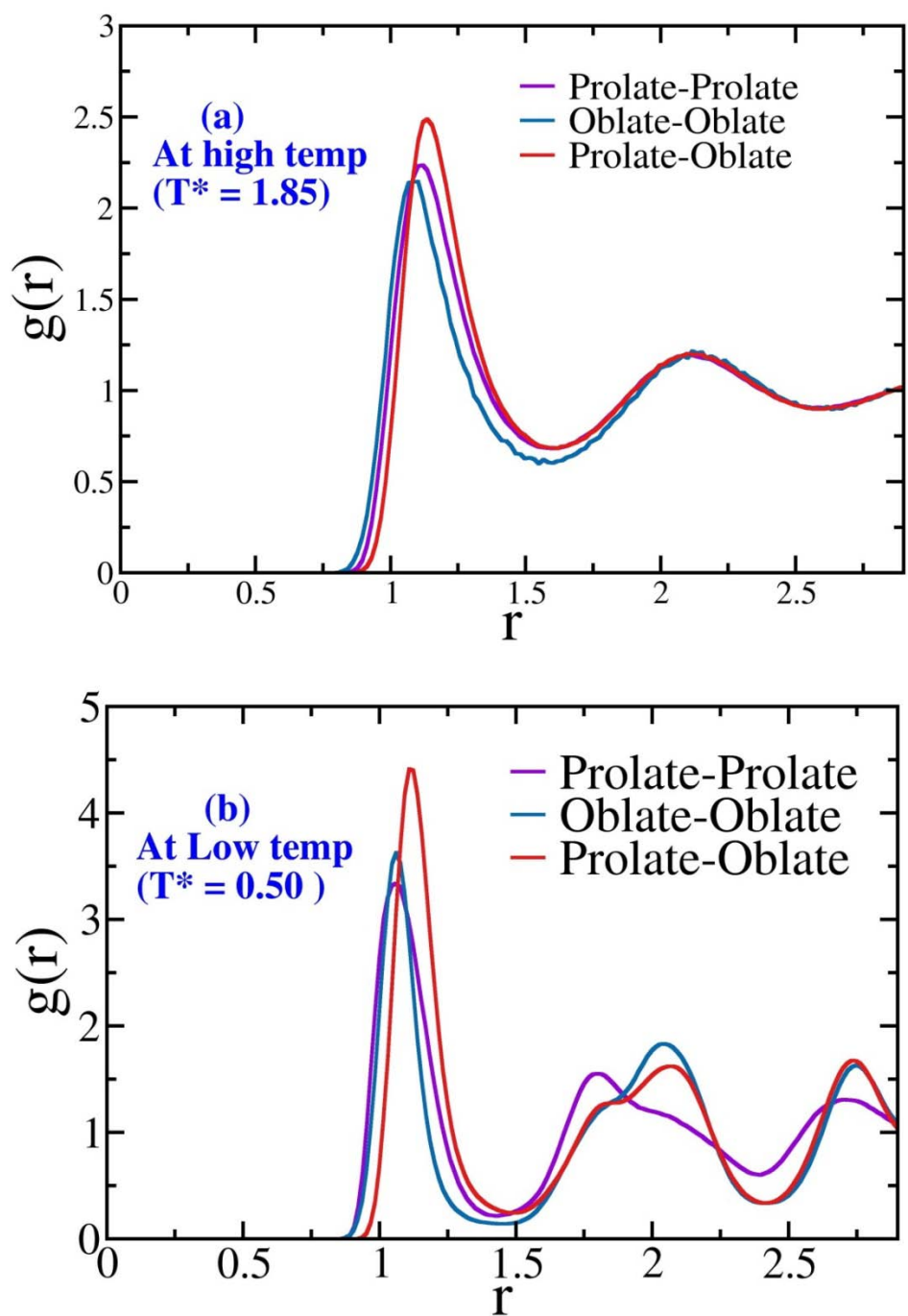

**Figure 8:** Plot of the partial radial distribution function of (a) the prolate-oblate binary mixture liquid at high-temperature T$^*$=5.00 and (b) the same liquid at the low-temperature T$^*$=0.50; at reduced pressure P$^*$=30. Note that the peak of Prolate-Prolate shows a minimum in comparison with that of Oblate-Oblate and

**Prolate-Oblate. Fig. 8(b) is the characteristic of glass transition as indicated by the splitting of the second peaks.**

In the binary liquid, particularly at higher temperatures, the entropic contribution keeps the binary mixture to be in homogeneous phase, as shown in Fig. **8(a)**. In the lower temperature (T* = 0.50), the prolate-prolate correlation gets substantially reduced while the prolate-oblate correlation gets enhanced. Figure **8(b)** shows that not only the large sharpening of the first peak of the radial distribution functions of prolate-oblate and oblate-oblate are found, but these correlation functions are also characterized by a split double peak, which is, of course, the signature of glass transition.[41] Similar split double peak is also seen in prolate-prolate, although the height is comparatively less in this case.

## V. CONCLUSION

The translational and rotational diffusion of tagged prolate in a sea of LJ spheres was studied in our earlier simulations. The present work could be regarded as the continuation of our earlier works. Here we have presented molecular dynamics simulations of oblates in a sea of prolates. In contrast to earlier studies, here we have employed the efficient software package LAMMPS that allowed us to consider large system sizes with longer MD simulation runs. The *main result of the present work is the observation of a breakdown of both the stick and slip hydrodynamic b.c. predictions to reproduce both rotational and translational friction values, for both prolates and oblates.* The marked breakdown of the slip b.c. of oblates and prolates is particularly striking as the common wisdom suggests that the said b.c. should capture rotational diffusion of ellipsoidal molecules. We note here that for spherical molecules, slip hydrodynamic b.c. is found to be a good predictor of the translational diffusion coefficient. For example, hydrodynamics predicts the values of translational diffusion of water molecules in neat water fairly accurately.

Another new aspect of this work is the inclusion of the orientational part in the interaction potential that is essential to describe molecular glasses and which is absent in the Kob-Andersen class of models. Calculated orientational time correlation function exhibits a faster decay for the oblates which is also a new result of this study. On the other hand, the translational diffusion of oblates is found to be slower than that of prolates. The anisotropy in the translational motion of oblates persists for a long time as compared to that of prolates.

One aspect of hydrodynamic prediction, however, has been found to hold. As D is inversely proportional to viscosity and rotational correlation time directly proportional, the product $D\tau_2$ is expected to be independent of temperature even when the shear viscosity increases by many orders of magnitude, as predicted by the combination of the Debye-Stokes-Einstein (DSE) and Stokes-Einstein (SE) equations. **Figure 5(a)** and **5(b)** indeed display such constancy of the product over a wide range of temperatures. for prolate and oblate, respectively.

We note that together with translational diffusion coefficient (D) and rotational time constant $(\tau_2)$ for both prolates and oblates, at multiple temperatures, we do have a substantial amount of quantitative data available with us. We have additionally studied the same composition and the same parameters at another pressure P*=20. The results on the validity of hydrodynamics remain the same.

The failures of both slip and stick hydrodynamic boundary conditions to describe the values of both translational diffusion coefficient and the rotational time constant is somewhat surprising.

While we do not fully know the origin yet, one possible reason is the strongly interacting limit used in this work where prolates and oblates interact strongly; this might lead to a scenario where this pair moves together. This shall lead to the breakdown of the slip boundary condition. In a subsequent work, we plan to study in detail the pair dynamics to understand the reason for the complete breakdown of hydrodynamic b.c. observed here. Another possible origin could be translation-rotation coupling affecting the dynamics not captured by hydrodynamic theories.

It is also worthwhile to explore the energy landscape manifestation in the view that the dynamical heterogeneity plays an important role in the decoupling between translational-rotational diffusion. Such decoupling could of interest as we approach the glass transition temperature. Work in this direction is under progress.

**Acknowledgments**

It is pleasure to thank Dr. Rajib Biswas (Asst. Professor, IIT Tirupati) for help and discussions. We thank the Department of Science and Technology (DST) and DST Nano Mission, India for partial financial support of this work. SS thanks DST for providing Post-Doctoral research fellowship and TS thanks CSIR-UGC for his research fellowship. BB thanks Sir J. C. Bose fellowship for partial support.